\documentclass{llncs}
\usepackage{makeidx}  

\usepackage{graphicx}

\usepackage[font=footnotesize]{subfig}
\usepackage{stfloats}

\usepackage{color}
\usepackage{url}

\newcommand{\URL}[1]{{\color{blue}\url{#1}}}

\usepackage{geometry}
\newcommand{\CHIPS}{{\emph{CHIPS}}}

\newcommand{\feeds}{{\emph{feeds}}}

\usepackage{enumitem}
\usepackage{graphicx}

\usepackage{wrapfig}

\begin{document}

\pagestyle{headings}  

\title{\vspace{0.25in} \emph{CHIPS} -- A Service for Collecting, Organizing, Processing, and Sharing Medical Image Data in the Cloud} 

\author{Rudolph Pienaar\inst{1,2} \and Ata Turk\inst{3} \and Jorge Bernal-Rusiel\inst{1} \and Nicolas Rannou\inst{5} \and Daniel Haehn\inst{4} \and P. Ellen Grant\inst{1,2} \and Orran Krieger\inst{3}}

\titlerunning{CHIPS: Cloud-based Mecial Image Processing}  
%
%
\authorrunning{Rudolph Pienaar et al.} 
\institute{Boston Children's Hospital, Boston, MA, 02115 USA,\\
\email{rudolph.pienaar@childrens.harvard.edu},
\and
Harvard Medical School, Boston, MA 02115, USA,
\and
Boston University, Boston, MA 02115, USA,
\and
Harvard School of Engineering and Applied Sciences, \\
Cambridge, MA 02138, USA,
\and
Eunate Technology S.L., Sopela, Spain
}

\maketitle              

\begin{abstract}
Web browsers are increasingly used as middleware platforms offering a central access point for service provision. Using backend containerization, RESTful APIs, and distributed computing allows for complex systems to be realized that address the needs of modern compute intense environments. In this paper, we present a web-based  medical image data and information management software platform called \CHIPS{} (\emph{C}loud \emph{H}ealthcare \emph{I}mage \emph{P}rocessing \emph{S}ervice). This cloud-based services allows for authenticated and secure retrieval of medical image data from resources typically found in hospitals, organizes and presents information in a modern feed-like interface, provides access to a growing library of plugins that process these data, allows for easy data sharing between users and provides powerful 3D visualization and real-time collaboration. Image processing is orchestrated across additional cloud-based resources using containerization technologies.

\end{abstract}

\section{Introduction}

Modern web browsers are becoming powerful platforms for advanced application development \cite{Hahn2012b,Ginsburg2011}. New advances in core web application technologies such as the modern web browsers' universal support of ECMAScript 5 (and 6) \cite{khan2014using}, CSS3 and HTML5 APIs have made it much more feasible to implement powerful middle-ware platforms for data management and powerful graphical rendering, as well as real-time communication purely in client-side JavaScript \cite{mwalongo2016state,bernal2017reusable}. The last decade has seen a slow, but steady, shift to fully distributed solutions using web-standards \cite{eckersley2003neuroscience,millan2014open,sherif2014cbrain,wood2014harnessing}, closely tracked by expressiveness of the JavaScript programming language. Web-based solutions are especially appealing as they do not require the installation of any client-side software other than a standard web browser which enhances accessibility and usability.

Unrelated to rise of web-technologies, a new emerging trend is the rapid adoption of containerization technologies. These have enabled the concept of \emph{compute} portability in a similar sense to \emph{data} portability. Just as data can be moved from place to place, containerization allows for operations on that data to also be moved from place to place.

To our knowledge, no web-based platform currently exists that provides data \emph{and} compute agnostic services (some services, such as CBRAIN \cite{sherif2014cbrain} and LONI~\cite{Rex:2003} provide conceptually similar approaches, but do not have deep connectivity to typical hospital database repositories), in particular collection, management, and real-time sharing of medical data, as well as access to pipelines that process that data. In this paper, we introduce \CHIPS{} (\textbf{C}loud \textbf{H}ealthcare \textbf{I}mage \textbf{P}rocessing \textbf{S}ervice). \CHIPS{} is a novel web-based medical data storage and data processing workflow service that provides strict data security while also facilitating secure, real-time interactive collaboration over the Internet and internal Intranets. 

\CHIPS{} is able to seamlessly collect data from typical sources found in hospitals (such as Picture Archive and Communications Systems, PACS) and easily export to approved cloud storage. \CHIPS{} not only manages data collection and organization, but it also provides a large (and expanding) library of pipelines to analyze imported data, and the containerized compute can execute in a large variety of remote resources. \CHIPS{} provides for persistent record and management of activity in \feeds{} as well as for powerful visualization of data. In particular, it makes use of the popular {\tt XTK} toolkit which was also developed by our team at the Fetal-Neonatal Neuroimaging and Developmental Science Center, Boston Children’s Hospital\footnote{\URL{http://fnndsc.babymri.org}} for the in-browser rendering and visualization of medical image data and can be freely downloaded from the web\footnote{\URL{http://goxtk.com}} \cite{haehn2014neuroimaging}. 

\section{Architectural Overview}

\subsection{Scope}

\begin{wrapfigure}{l}{6.5cm}
    \centering
    \includegraphics[width=6cm]{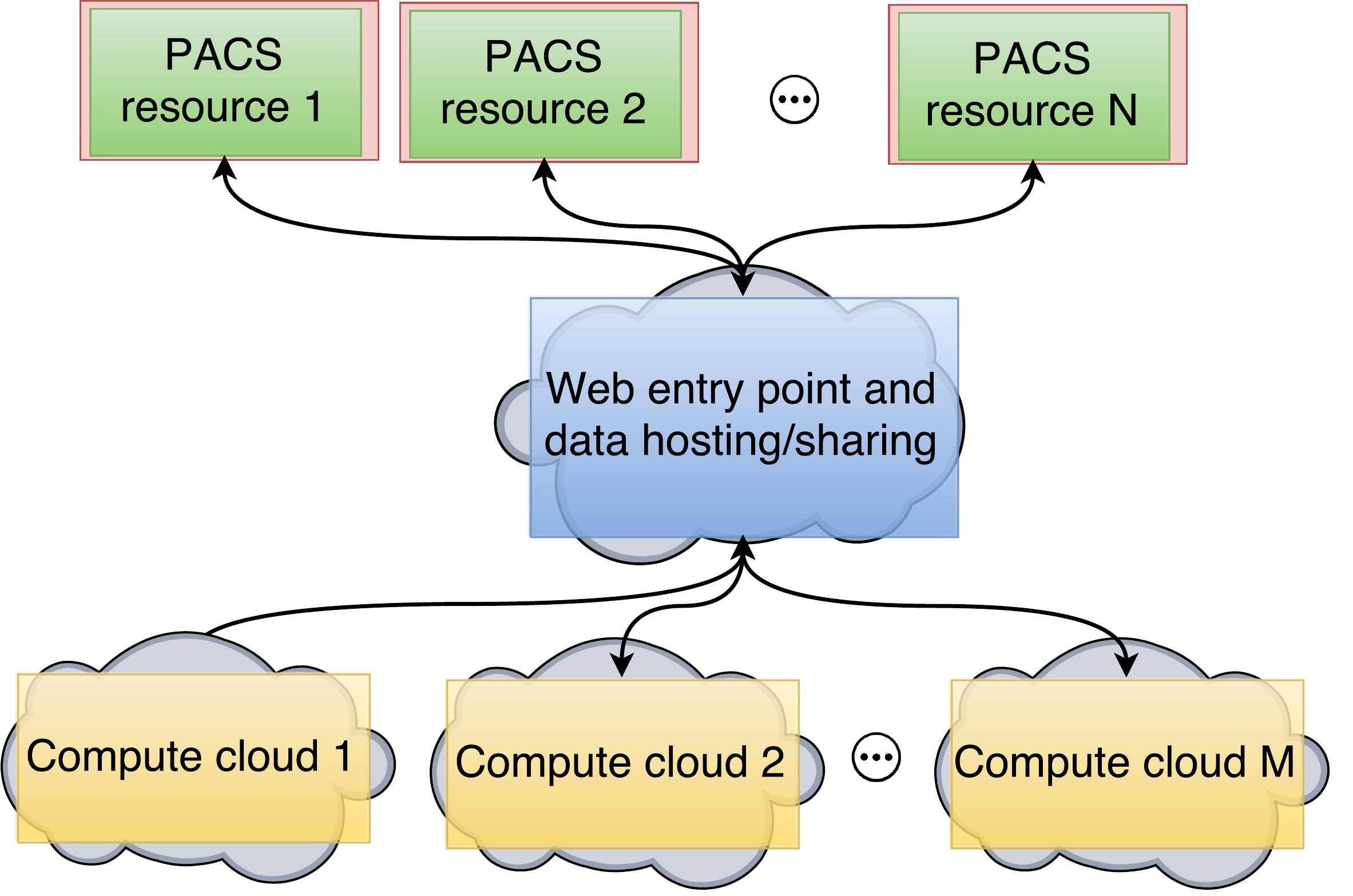}
    \caption{\label{fig:chris_expand} \CHIPS{} connects multiple input PACS sources to multiple ``cloud" compute nodes.}
    \noindent \hrulefill
    \vspace{-30pt}
\end{wrapfigure} 

The creation of \CHIPS{} has been motivated by both clinical and research needs. On the clinical side, \CHIPS{} was built to provide clinicians with easy access to large amounts of data (especially from hospital image databases like Picture Archive and Communications Systems -- PACS), to provide for powerful collaboration, and to allow for easy access to a library of analysis processes or pipelines. On the research side, \CHIPS{} was designed to allow computational researchers to test and develop new algorithms for image processing across heterogeneous platforms, while allowing life science researchers to focus on their research protocols and data processing, without needing to spend time on the minutiae of performing data analysis.


The system design is highly distributed, as shown in Figure \ref{fig:chris_expand}, which shows a \CHIPS{} deployment connected to multiple input sources and multiple compute sources. Though the figure suggests a single, discrete central point, components of \CHIPS{} do reside on each input (PACS) and compute location.

\subsection{Distributed Component Design}

\begin{wrapfigure}{r}{6.5cm}
    \centering
    \vspace{-40pt}
    \includegraphics[width=6cm]{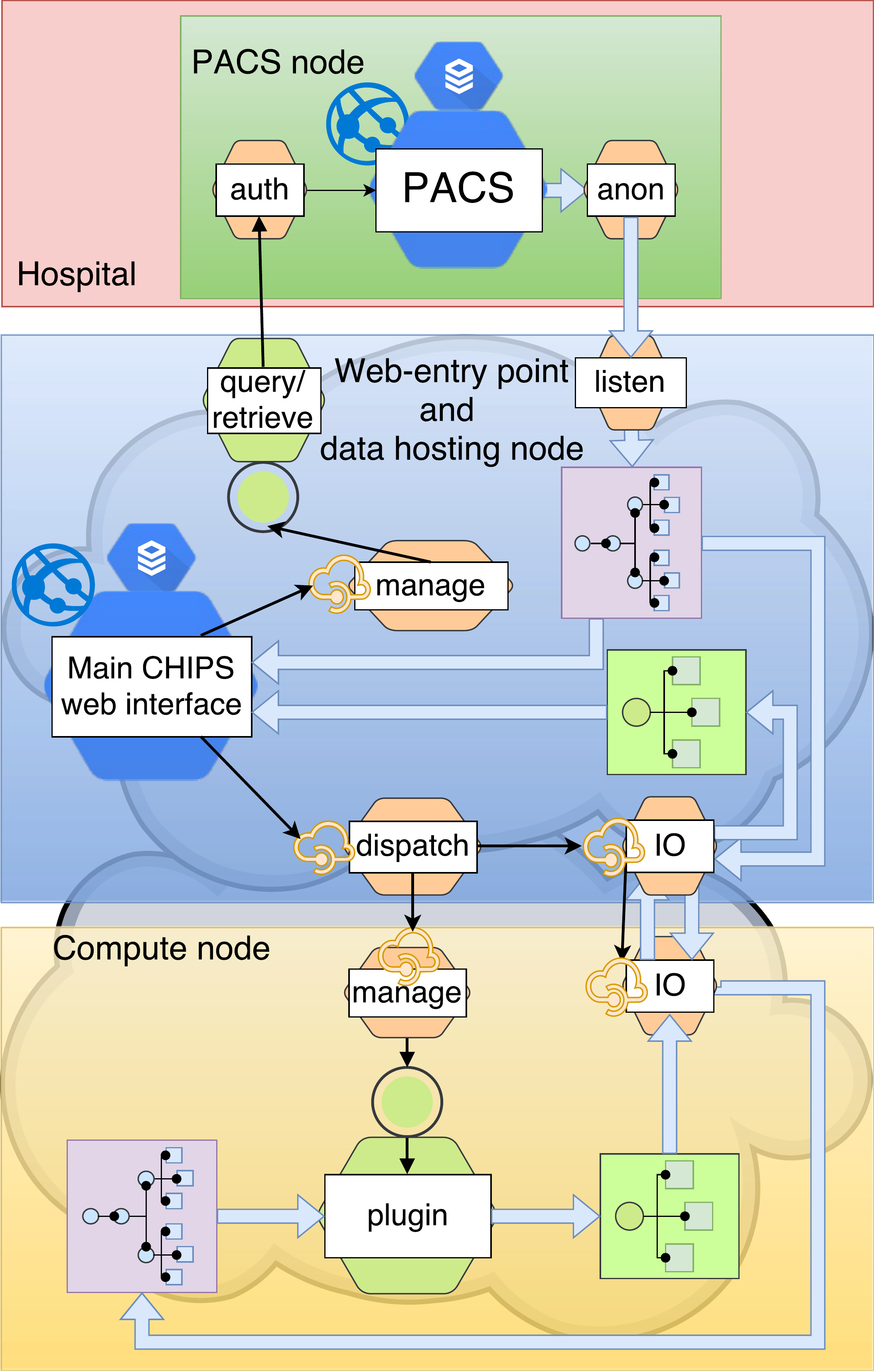}
    \caption{\label{fig:chris_distributed_workflow} \small The internal \CHIPS{} logical architecture.}
    \noindent \hrulefill
    \vspace{-20pt}
\end{wrapfigure} 

Architecturally \CHIPS{} is not a single monolithic system, but a distributed collection of interconnected components, including a front-end webserver and web-based UI; a core RESTful back-end central server that provides access to all data, feeds, users, etc; a DICOM/PACS interface; a set of independent RESTful microservices that handle inter-network data IO and also remote process management, and a core cloud-based computational platform that orchestrates offloading of image processing pipelines to some remote cloud-based compute -- see Figure \ref{fig:chris_distributed_workflow}.

The top the red box of Figure \ref{fig:chris_distributed_workflow} contains the \emph{PACS node} and represents the Hospital image data repository. The second blue box, labeled {\it Web-entry point and data hosting node} contains the main \CHIPS{}  backend and is presented as being in a ``cloud" (i.e. some resource that is accessible from the Internet). Finally, the bottom yellow box is shown on a separate ``cloud" to emphasize that it is topologically distinct from the {\it Web-entry point}. 

The logical relationships between data (represented as the rectangles with a tree structure) and compute elements denoted by the named hexagons is shown by either data connectors (thick blue arrows) or control connections (single line arrows). In the syntax of the diagram, the stylized cloud icon touching some of the boxes denotes that these compute elements are controlled by a REST API, while the sphere icon denotes web-access. 

An remote compute is denoted by {\tt plugin}, which is controlled by a {\tt manage} component. In the most abstract sense, the {\tt plugin} processes an input data structure, and outputs a transformed data structure (the two tree graphs as shown). File transfer between the data cloud and compute cloud is performed by the file {\tt IO} handler component. A {\tt query/retrieve} process in the data cloud connects to an authentication process, {\tt auth} in the Hospital network, while on-the-fly anonymization of DICOM images is handled by process anonymizer {\tt anon}. Finally the {\tt dispatcher} is a component that determines what compute node (or cloud) is best suited for the data analysis at hand. The circle icon attached to the {\tt manage} and {\tt plugin} icons implies the attached process and can provide real-time feedback information to other software agents about the controlled process via its own REST interface.

\subsection{Pervasive containerization}

\CHIPS{} is designed as a distributed system, and the underlying components are containerized (currently using docker\footnote{\URL{https://www.docker.com}}. In Figure \ref{fig:chris_distributed_workflow}, the {\it Main CHIPS web interface} and associated backend database is housed within a single container\footnote{\URL{https://github.com/FNNDSC/ChRIS_ultron_backEnd}}. Input data and processed results are accessible in the hosting node and volume mapped as appropriate to this back end.
Other components of \CHIPS{} in the web-entry node are similarly containerized. This includes the {\tt manage}\footnote{\URL{https://github.com/FNNDSC/pman}} block, which is responsible for spawning processes on the underlying system. Not only does {\tt manage} provide the means to start and stop processes, but it also tracks the execution state, termination state, and standard output/error streams of the process. The {\tt manage} component has a REST interface through which clients can start/stop and query processes. 

Also containerized is the {\tt IO}\footnote{\URL{https://github.com/FNNDSC/pfioh}} component that can transfer entire directory trees across network boundaries from one system to another as well as the {\tt dispatch}\footnote{\URL{https://github.com/FNNDSC/swarm}} component that can orchestrate multiple processing jobs as handled by {\tt manage}. The {\tt plugin} container houses the particular compute to perform on a given set of data, and is spawned by the {\tt manage} component under direction of the {\tt dispatch}. Since the compute typically occurs on a separate system to the data hosting node, the {\tt IO} containers perform the necessary transmit of data to this compute system, as well as the retrieve of resultant data back to the data node, allowing the web container to present (and visualize) results to the user.

\section{UI Considerations}


\begin{wrapfigure}{l}{7.5cm}
    \centering
    \vspace{-30pt}
    \includegraphics[width=6cm]{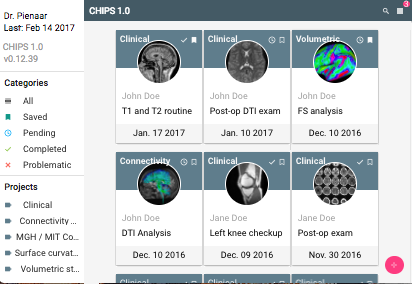}
    \caption{\CHIPS{} home page with a ``cards" organization.}
    \noindent \hrulefill
    \label{fig:ChIPS}
    \vspace{-10pt}
\end{wrapfigure} 

Figure \ref{fig:ChIPS} shows the home page view on first logging into the system. Studies that have been ``sent" to \CHIPS{} appear in their own ``cards" on the user's home page with a small visualization of a represented image set of the study. Various control on this home page allow users to organize/tag ``cards" in specific projects (or folders), remove cards, bookmark for easy access, etc. New cards can be generated by clicking on the \textcircled{+} icon and choosing an activity (such as PACS Query/Retrieve), and any card can be seamlessly shared with other users of the system.

\begin{wrapfigure}{r}{7.5cm}
    \centering
    \vspace{-20pt}
    \includegraphics[width=6cm]{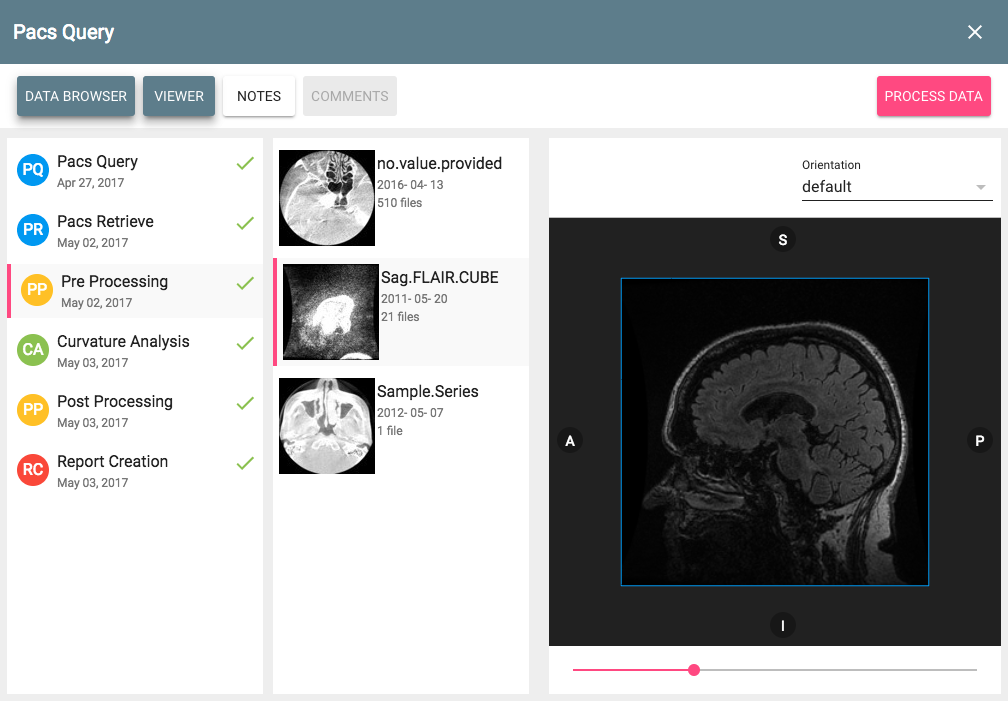}
    \caption{\label{fig:feed-viewer} \small Visualizing pulled and processed data.}
    \noindent \hrulefill
    \vspace{-50pt}
\end{wrapfigure}

On selecting a given feed, the core image data in that feed is visualized in a rich, web-based viewer -- see Figure \ref{fig:feed-viewer}. Various tabs and elements of the feed view provide different perspectives on the data, and also provide the ability to annotate notes, or add comments. As in the feed view, a \textcircled{+} icon is also present, and if selected, opens a ribbon of ``plugins" (or ``apps") to run on the data contained in the feed. For example, certain plugins might perform a surface reconstruction of the brain surface with tissue segmentation (for example, a FreeSurfer plugin).

The interface semantics within a feed are straightforward: a user clicks on the feed and enters the top level data view. Once a plugin from the \textcircled{+} is applied, the feed data is processed accordingly. When the plugin is completed, its output files are also organized in the feed in a logical tree view (accessible via the left "Data" tab) in a manner akin to an email thread. In this manner, the thread of execution from data $\rightarrow$ plugin $\rightarrow$ data is defined -- in effect building a workflow.

Any image visualized can also be shared in real-time using collaboration features built into the viewer library and leveraging the Google Drive API and Google Realtime API \cite{bernal2017reusable}.

\section{Big Data Infrastructure}

\begin{wrapfigure}{l}{10.5cm}
    \centering
    \includegraphics[width=10cm]{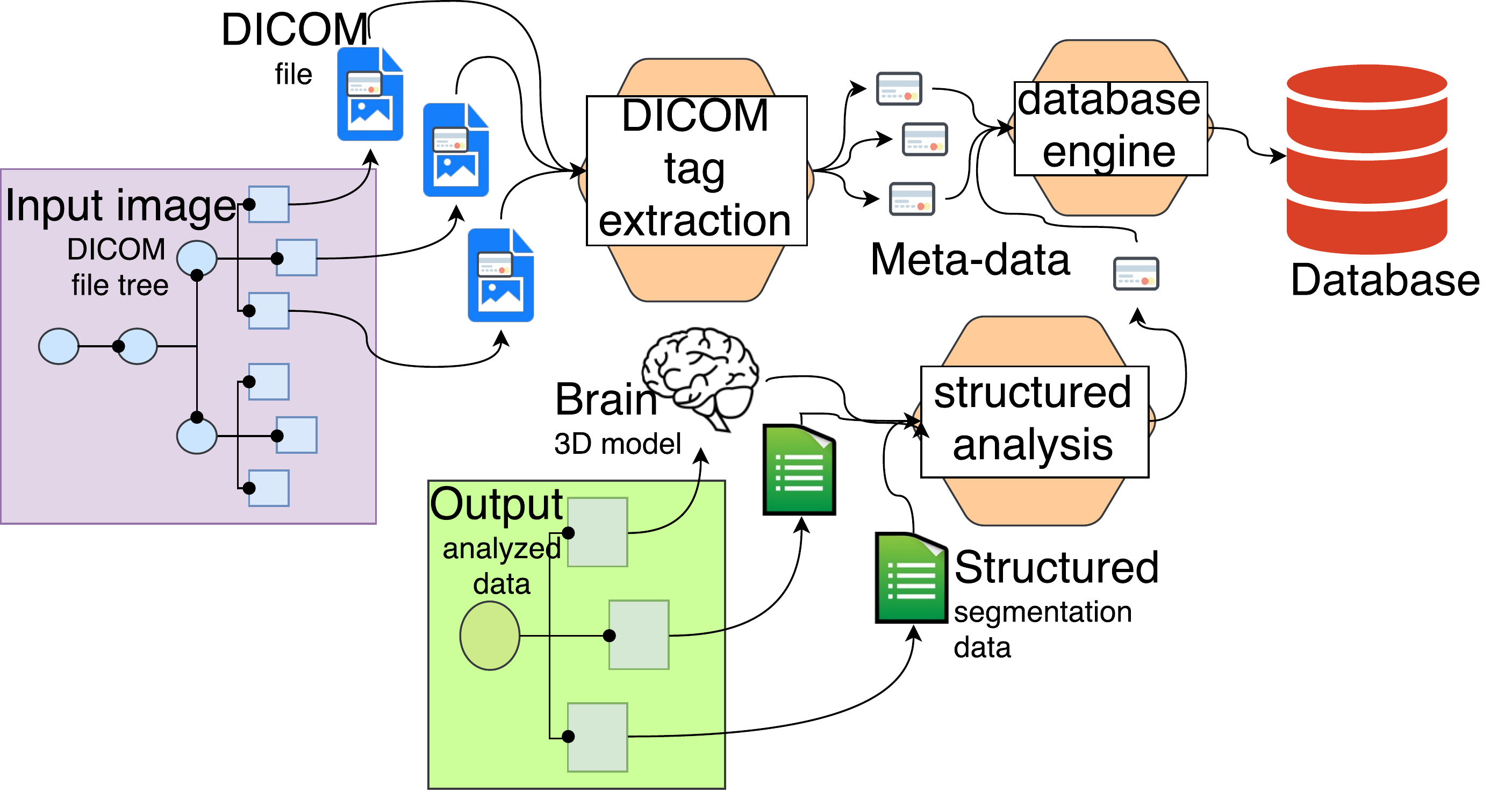}
    \caption{\label{fig:bigData} \small Big data pre-processing.}
    \noindent \hrulefill
    \vspace{-10pt}
\end{wrapfigure} 

An important component of \CHIPS{} lies in creating a foundation suitable for future support of ``data mining". Recently, the term \emph{Big Data} has come into common parlance, especially in the context of informatics \cite{Provost2013,Swan2013,JCP:JCP24662}. Despite the term and the use of \emph{Big}, the concept often refers to the use of predictive analytics and other advanced data analytics tools that extract meaning from sets of data and does not necessarily to the particular size of the data set.

In healthcare, big data analytics has impacted the field in very specific areas such as clinical risk intervention, waste and care variability reduction, and automated reporting. However, as a field, biomedical imaging has not especially benefited from big data approaches due to the unstructured nature of image data, complexity of results from analysis in terms of data formats (again usually unstructured), simple quality issues such as noise in image acquisitions, etc.

\CHIPS{} constructs a framework to allow big data methods to be used in this image space. Consider that the incoming source data to \CHIPS{} are DICOM images that by their nature contain a large amount of meta information, most of which is non PHI and will be left unchanged by the anonymization processes. Information about the scanning/imaging protocol, acquisition parameters, as well as certain non-PHI demographics such as patient sex and age can be meaningfully databased. Moreover, the application of an analysis pipeline to an image data-set can in turn result in large amounts of meaningful data that can be databased and associated with the incoming source data. For example, FreeSurfer, which is dockerized as a plugin in the \CHIPS{} system produces volumetric segmentations and surface reconstructions on raw input MRI T1 weighted data \cite{Dale99corticalsurface-based,Fischl99corticalsurface-based,freesurfer}. 

In Figure \ref{fig:bigData} input raw DICOM (purple block) and output processed data from the DICOMs (green block) are shown. A {\tt DICOM tag extraction} process removes the image meta data and associates this information with the particular image record. DICOM data is regularly formatted and easily extracted. Importantly, for the output data, and assuming the output data is a 3D surface reconstruction and tables of brain parcellation volume values, a {\tt structured analysis} process regularizes all this information into meta data that will be added to the space of data pertaining to this image record. This processing will lay the ground work on which data analytics can explore and mine for relations between (for example) input acquisition parameters and pipeline output results, or simply mine across output results for hidden trends in data trajectories (for example volumetric changes with age or sex).


\section{Conclusion and Future Directions}

\CHIPS{} is a distributed system that provides a single, cloud-based, access point to a large family of services. These include: (a) accessing medical image data securely from participating institutions with authenticated access and built-in anonymization of collected image data; (b) organizing collected data in a modern UI that allows for easy data management and sharing; (c) performing processing on images by dispatching data to remote clouds and controlling/managing remote execution on these resources; (d) powerful real-time collaboration on images using secure third party services (such as the Google RealTime API); and intuitively constructing medical image processing workflows. \CHIPS{} is not only a medical data management system, but strives to improve the quality of healthcare by allowing clinical users the ability to easily perform value added processing and sharing of data and information. Current and future directions for \CHIPS{} include facilitating the construction of big-data frameworks and allowing for users to simply construct experiments for data analytics and various machine learning pipelines.

All analysis and development conducted by the \CHIPS{} system at the Boston Children's Hospital was conducted under relevant Institutional Review Board approval, which governed access to image data and controlled the scope of sharing of such data.

\bibliographystyle{splncs03}
\bibliography{pienaar-references}

\end{document}